\documentstyle[aaspp4,12pt,epsfig,color]{article}
\begin{document}

\title{Cloning Hubble Deep Fields II: Models for Evolution\\
by Bright Galaxy Image Transformation}

\author{Rychard Bouwens} 
\affil{Physics Department,
   University of California,
    Berkeley, CA 94720; bouwens@astro.berkeley.edu}
\author{Tom Broadhurst}
\affil{Astronomy Department,
    University of California,
    Berkeley, CA 94720; tjb@astro.berkeley.edu}
\centerline \&
\author {Joseph Silk}
\affil{Astronomy and Physics Departments, and Center for Particle
    Astrophysics, University of California,
     Berkeley, CA 94720; silk@astro.berkeley.edu}

\begin{abstract}
 
In a companion paper we outlined a methodology for generating
parameter-free, model-independent ``no-evolution'' fields of faint
galaxy images, demonstrating the need for significant evolution in the
HDF at faint magnitudes.  Here we incorporate evolution into our
procedure, by transforming the input bright galaxy images with
redshift, for comparison with the HDF at faint magnitudes.  Pure
luminosity evolution is explored assuming that galaxy surface
brightness evolves uniformly, at a rate chosen to reproduce the I-band
counts.  This form of evolution exacerbates the size discrepancy
identified by our no-evolution simulations, by increasing the area of
a galaxy visible to a fixed isophote.  Reasonable dwarf-augmented
models are unable to generate the count excess invoking moderate rates
of stellar evolution.  A plausible fit to the counts and sizes is
provided by `mass-conserving' density-evolution, consistent with
small-scale hierarchical growth, where the product of disk area and
space density is conserved with redshift.  Here the increased surface
brightness generated by stellar evolution is accomodated by the
reduced average galaxy size, for a wide range of geometries.  These
models are useful for assessing the limitations of the HDF images, by
calculating their rates of incompleteness and the degree of
over-counting. Finally we demonstrate the potential for improvement in
quantifying evolution at fainter magnitudes using the HST Advanced
Camera, with its superior UV and optical performance.

\end{abstract}

\keywords{galaxies: evolution --- galaxies: scale-lengths}

\section{Introduction}

Deep HST images have proven difficult to interpret.  One might have
imagined that clear pictures of the distant Universe would speak for
themselves, revealing directly how galaxies formed and evolved.
Instead, interpretation has been hampered by our ignorance of the UV
properties of local galaxies.  In a companion paper (Bouwens,
Broadhurst, \& Silk 1998; BBS-I), we showed how this problem can be
overcome using a redshift-complete sample of bright galaxies
contructed from the UV-optical HDF images and follow-up spectroscopy.
Clear evolutionary trends were identified by projecting this bright
sample to much fainter magnitudes in a purely empirical and
model-independent way.  In the present paper, we explore simple models
for evolution by modifying our procedure to generate realistic deep
fields for exploring the question of evolution to magnitudes far
fainter than accessed by spectroscopy.  We commence with nearby
galaxies and project them back into the past, modifying the number,
surface brightness and sizes of the images in ways designed to embody
both luminosity and density evolution and to explore the possible role
of dwarf galaxies.  These simple image transformation models represent
a heuristic alternative to other more complicated evolutionary
approaches with more model-dependent and parameter-laden
representations of galaxies at both low and high redshift.

At present this empirical modelling is still a more reliable guide to
the process of galaxy evolution than that derived from numerical work
in the context of the hierarchical models for the growth of structure.
High-resolution codes which simulate the gravitational interactions of
gas and cold dark matter and incorporate cooling following standard
atomic physics produce small, dense and rapidly rotating disks in the
center of massive haloes (Navarro \& White 1994; Navarro \& Steinmetz
1997).  Of course, ad-hoc heat input from supernovae can presumably be
added to achieve larger disks, but, nonetheless, even with this level
of freedom, the rotation curves of dwarf galaxies such as DDO154
(Carignan \& Freeman 1988) and other newly discovered dwarfs (Cote,
Freeman, \& Carignan 1997) defy simple explanation (Navarro, Frenk, \&
White 1996; Burkert \& Silk 1997).  Semi-analytic attempts to mock up
hierarchical evolution (cf. Baugh, Cole, \& Frenk 1996; Kauffmann,
Guiderdoni, \& White 1994), though providing interesting
interpretations through which to understand the various observables
relevant to galaxy formation and evolution, still include many free
parameters.  Improved modelling of the observations will undoubtedly
require a deeper understanding of the interplay of physical processes
relevant to star formation.

In \S1, we describe our models for evolution and compare them with the
HDF observations in \S2.  In \S3, we discuss the results, and in \S4
we present our conclusions and discuss future prospects.  As in BBS-I,
we adopt $H_0 = 50\,\textrm{km/s/Mpc}$ and express all magnitudes in
this paper in the AB\footnote{$m \textrm{(AB)} = -2.5 \log f_{\nu}
\textrm{(erg/cm/cm/s/Hz)} - 48.60$ (Oke 1974)} magnitude system
(defined in terms of a flat spectrum in frequency).  Also, to
associate the HDF bands with their more familiar optical counterparts,
we shall refer to the $F814W$, $F606W$, $F450W$, and $F300W$ bands as
$I_{814}$, $V_{606}$, $B_{450}$, and $U_{300}$, respectively,
throughout this paper.

\section{Standard Models}

We present simple augmentations to our procedure for generating
realistic deep fields as described in detail in BBS-I.  We incorporate
the usual phenomenological models proposed to explain the number
counts: pure luminosity evolution, ``mass-conserving'' density
evolution, and the contribution of an evolving low-luminosity (dwarf)
population, all of which have been claimed to be important for
matching the number counts at faint magnitudes.  We proceed by
performing simple scalings of the images of our bright HDF galaxy
sample (see BBS-I) in surface brightness, size, and the space density
in the simplest manner possible in order to achieve rough agreement
with the number counts in the $I_{814}$ band.

Unfortunately, by virtue of its size, our bright HDF sample lacks low
luminosity galaxies, and therefore dwarfs are necessarily included in
an ad-hoc manner, requiring that we specify in addition to their
evolution their image profiles and luminosity functions.  This
latitude has generated a large literature on their potential
contribution (Kron 1982; Cowie, Songaila, \& Hu 1991; Gronwall \& Koo
1995; Ferguson \& McGaugh 1995; Driver \& Phillips 1996).  In the
present work, we have chosen to consider a somewhat conservative
``maximal'' estimate of this contribution, using constraints on the
faint end slope of the LF from recent redshift surveys.

\subsection{Luminosity Evolution}

To obtain a rough idea of the effect of luminosity evolution on deep
field images, we modify our simulation procedure to include a simple
$(1+z)^b$ scaling in surface brightness, a very simple prescription
for obtaining fair agreement with the number counts in the $I_{814}$
band to some specified magnitude limit.  Of course, one could model
this form of evolution with more sophistication using the usual
assumptions (Pozzetti, Bruzual, \& Zamorani 1996; Ferguson \& Babul
1998), even implementing it on a pixel-by-pixel basis where the
observed pixel colours could be used to model the spatial history of
star formation.  However, we have decided not to pursue this here
because of the many additional assumptions required and the simple
fact that such models, as we will illustrate, would inevitably seem to
produce galactic populations with sizes that are too large.  In BBS-I,
such a trend toward small sizes was already evident.

\subsection{Density Evolution}

To estimate the manner in which density evolution would alter the
predictions of our empirical no-evolution method, for simplicity, we
have chosen to scale the number densities of each bright galaxy in the
sample as $(1+z)^Q$ without changing their relative proportions.  To
make this simple model sensible, we decrease the metric sizes subject
to the simple constraint that the integrated mass is preserved,
similar to the models proposed by Rocca-Volmerange \& Guiderdoni
(1990) and Broadhurst, Ellis, \& Glazebrook (1992).  For disk
galaxies, this simply translates into the requirement that galaxies
change in area at a rate, $(1+z)^{-Q}$, inversely proportional to the
space density evolution.  As discussed in Broadhurst et al.\ (1992),
it is also necessary in the context of this simple ``merger''
prescription to add luminosity evolution, since otherwise the number
counts are virtually unaffected.  We also parameterize this luminosity
evolution as a $(1+z)^b$ scaling.

We shall avoid consideration here of various arguments against this
scenario, for which high rates of merging are claimed to be
problematic (T\'oth \& Ostriker 1992, Lacey \& Cole 1993; Dalcanton
1993; Roukema \& Yoshii 1993), recognising that from a purely
empirical perspective this model has considerable intrinsic appeal.

For both the luminosity evolution and density evolution models, we
have listed the best-fit parameters in Table 1.  Except for simple
scalings in number density, size, and surface brightness, we performed
these simulations in an identical manner to that described in BBS-I
using the Coleman, Wu, \& Weedman (1980) SED templates.  Note that we
self-consistently derived $V_{max}$ and the number densities for each
geometry and evolutionary scenario.  For the purposes of illustration,
we present our pure luminosity simulations in Figure 1 for three
different geometries ($\Lambda = 0.9$/$\Omega = 0.1$; $\Omega = 0.1$;
$\Omega = 1$) and an HDF density evolution simulation in Figure 2.

\begin{figure}
\begin{center}
\resizebox{11.5cm}{!}{\includegraphics*[125,110][450,620]{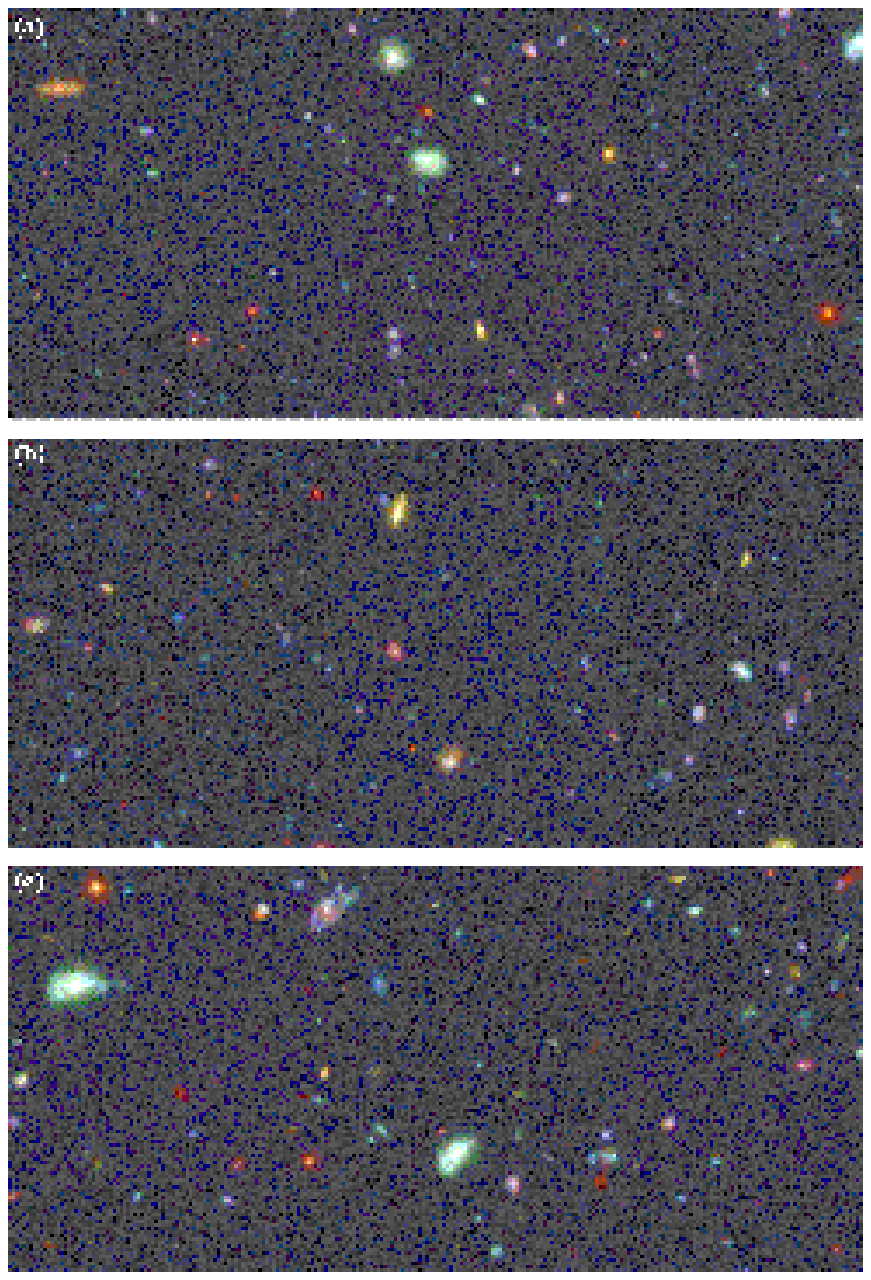}}
\end{center}
\caption{ Panel (a) shows a simulated 96'' x 46'' colour image
generated from the $B_{450}$, $V_{606}$, and $I_{814}$ bands for our
$\Omega = 0.1$/$\Lambda = 0.9$ LE prescription with $\Omega = 1$
constructed with pixel size, signal-to-noise, and PSF identical to
that of the HDF.  Panels (b) and (c) is similar to panel (a), except
using the $\Omega = 0.1$ and $\Omega = 1.0$ LE prescriptions,
respectively.  All LE models produce objects which are larger on
average than the data.  Also note prominent examples of objects
breaking-up into HII regions at high redshift apparent in several
panels.\label{ple}}
\end{figure}

\begin{figure}
\begin{center}
\resizebox{11.5cm}{!}{\includegraphics*[125,130][450,600]{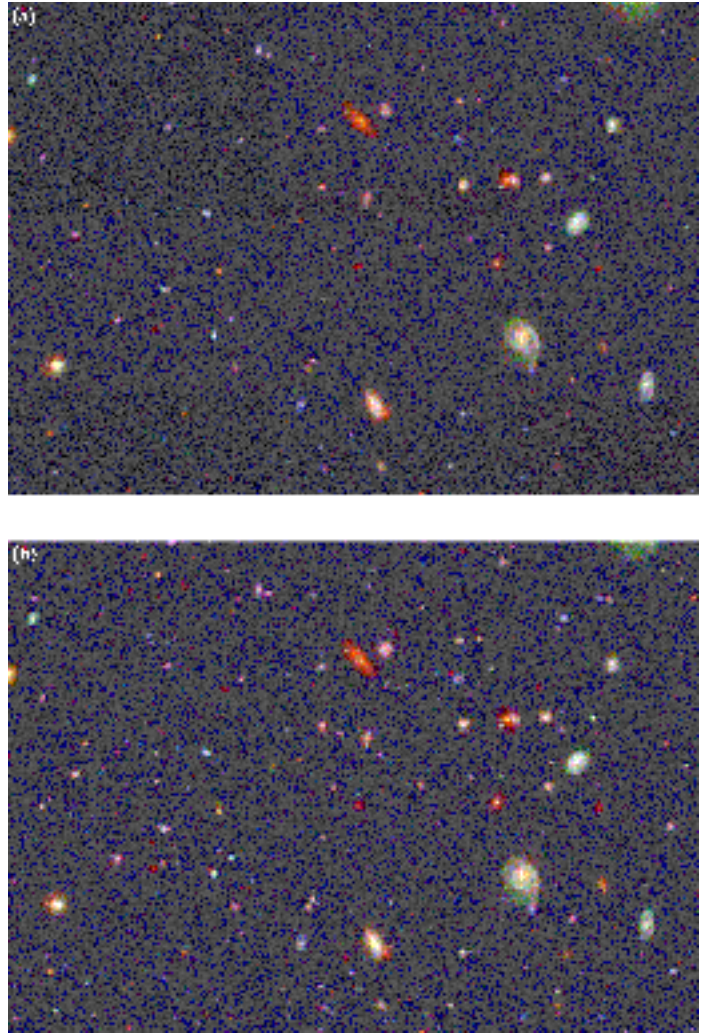}}
\end{center}
\figcaption{Panel (a) shows a simulated 100'' x 64'' colour image
generated from the $B_{450}$, $V_{606}$, and $I_{814}$ bands for our
merging prescription where $L \propto (1+z)^1$ and $n \propto (1+z)^2$
in a $\Omega = 1.0$ geometry, constructed with pixel size,
signal-to-noise, and PSF identical to that of the HDF.  Both the
number counts and angular size distributions are in good agreement
with the HDF.  Panel (b) shows a similarly-sized region of the same
model, except that the sensitivities of the Advanced Camera are used,
here estimated to be 5, 4, and 5 times larger than for WFPC2 in the
$B$, $V$, and $I$ bands, respectively.  Notice the large increase in
the number of faint galaxies expected.\label{ac}}
\end{figure}

\placetable{LMparam}
\placefigure{ple}
\placefigure{ac}

\subsection{Low-Luminosity Galaxies}

Currently, our input bright galaxy sample does not extend faintward of
$M_{b_j}=-18$.  Consequently, we have included a model low-luminosity
population to explore the potential role of a dwarf population.  Such
a model is particularly important to explore since it is well
understood that in the absence of evolution, the low-luminosity
galaxies comprise an increasing contribution to the counts at faint
magnitudes by virtue of their relatively small k-corrections and the
increasing volume available to these galaxies relative to the more
luminous galaxies, particularly for large $\Omega$ (Kron 1982).
  
Given the latitude of possible models permissible for the low
luminosity galaxies, we shall simply use the same luminosity function,
initial mass functions (IMF) and star formation histories as are given
in the Pozzetti et al.\ (1996) pure luminosity evolution model for the
Sabc and Sdm spectral types, except that we shall adopt a
normalization which is 50\% higher than the value prescribed in
Pozzetti et al.\ (1996) and we truncate the luminosity function at
absolute magnitudes $b_j < -18$, brightward of which our bright HDF
sample (see BBS-I) is well-represented.  Our low-luminosity population
already overproduces the low-redshift galaxies in the CFRS, but might
be tolerated depending on where this survey's 19\% incompleteness lies
in redshift.  These dwarfs are all endowed with exponential profiles
with a $b_j=22.65\, \textrm{mag/arcsec}^2$ central surface brightness.
We have intentionally chosen a central surface brightness lower than
the commonly observed central surface brightness of 21.65
$b_j\,\textrm{mag/arsec}^2$ typical of high luminosity objects, as a
way to conservatively account for the claimed correlation between
surface brightness and luminosity (McGaugh \& de Blok 1997).  A
summary of the parameters used in this model is provided in Table 2.

\placetable{lowl}

Although there continues to be debate about the importance of low
luminosity galaxies, both locally and in the past, we feel that our
treatment of dwarfs is generous, pushing the limits of what is
supported by observation.  Relative to other determinations of the
local field galaxy luminosity function, we have a fairly high
normalization and a relatively steep faint-end slope $-1.24$, much
like the LF claimed by the fairly deep ESP survey of Zucca et al.\
(1997) and steeper than the well defined APM luminosity function of
Loveday et al.\ (1992).  At fainter magnitudes, no redshift survey has
ever shown as many low redshift objects as one would expect if the
density of dwarfs was much larger than this (Broadhurst, Ellis, \&
Shanks 1988; Glazebrook et al.\ 1995a; Cowie et al.\ 1996; Lilly et
al.\ 1995; Ellis et al.\ 1996; Heyl et al.\ 1997).  Indeed, with
regard to the observed distribution of redshifts from the CFRS, our
generous model for the low-luminosity population of galaxies already
generates a $\sim50\%$ excess at the low redshift end.  Despite these
constraints, a larger population of dwarfs could in principle be
admitted within these observational constraints by adopting a much
steeper slope to the luminosity function, a possibility for which
there has been some recent support (Loveday 1997).

For this model, we have generated Monte-Carlo catalogues and simulated
images in $U_{300}$, $B_{450}$, $V_{606}$, and $I_{814}$, equal in
area to 4 times that of the HDF.  The dwarfs are placed on the image
at random positions and inclination angles, assuming no extinction and
smoothed with an unsaturated, relatively isolated stellar PSF taken
from the HDF.  Then both poissonian and sky noise are placed on the
images, after applying the noise kernel to reproduce the drizzled
properties of the noise.  We calculate the colours for these galaxies
with their chosen star-formation histories using a recent version of
the Bruzual \& Charlot spectral synthesis tables, compiled in
Leitherer et al.\ (1996) (see Charlot, Worthey, \& Bressan 1996 for a
description of these type of models.)  We recover objects off the
resulting dwarf-augmented images using SExtractor in exactly the same
manner that we recover objects from the HDF.

\section{Results}

\subsection{Number Counts in the $I_{814}$ band}

As in BBS-I, we begin by examining the number counts derived from the
above models in the $I_{814}$ band due to relatively small
uncertainties inherent in determining the fluxes each of the input
galaxies would have to $z\sim3$ in this band.  The pure bolometric
luminosity evolution results are presented in Figure 3 as a dashed
line.  The rate, parameterized as $(1+z)^b$, ranges over $b=1.4-2.5$
depending on the geometry, being smaller for models with larger volume
elements.  At the faintest magnitudes, the model counts fall off with
respect to the data, a feature more the result of a higher level of
incompleteness than a lack of volume at high redshift.  Incompleteness
sets in because the higher mean redshifts generated with this form of
evolution results in a lower mean surface brightness at high redshift,
where the rate of cosmological dimming $(1+z)^4$ is much greater than
the compensation from luminosity evolution.  Note that these rates of
luminosity evolution are roughly consistent, albeit a little lower
than the best fit found by Lilly et al.\ (1998) of $(1+z)^{2.7}$ for
the spiral galaxy population ($\Omega = 1$, $0\lesssim z \lesssim 1$).

\begin{figure}
\epsscale{0.9}
\plotone{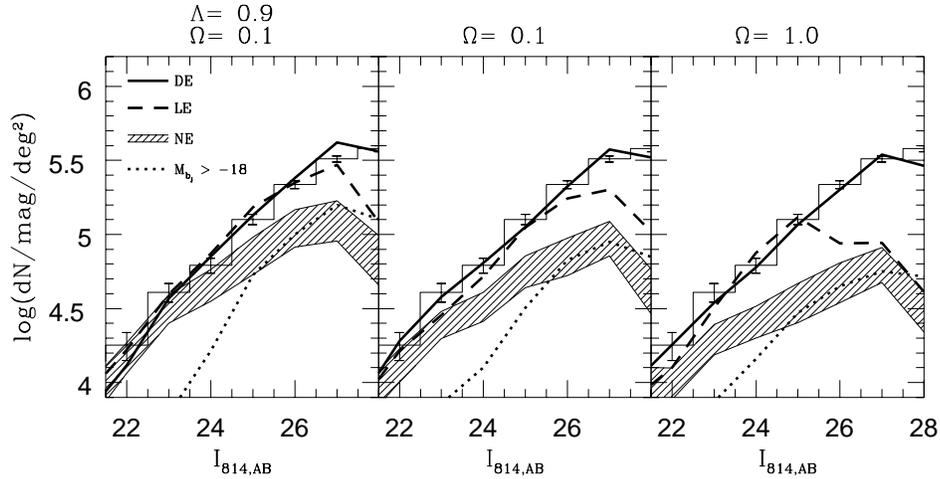}
\caption{A comparison of the observed $I_{814,AB}$-band number counts
(histogram with $1 \sigma$ Poisson errors) with those recovered from
our simple merging prescription (solid line) and our pure luminosity
prescription (dashed line).  For comparison, the counts recovered from
the no-evolution simulations are shown (the hatched area representing
the estimated 1 $\sigma$ range in these counts based on the finite
size of our bright input sample) along with those estimated to derive
from low-luminosity galaxies (dotted line).  All cases are shown for
$\Omega = 0.1$/$\Lambda = 0.9$, $\Omega = 0.1$, and $\Omega = 1$
geometries.  Note that the turnover in the number counts at faint
magnitudes ($I_{814,AB} \sim 26$) is more a result of incompleteness
than a lack of volume at high redshifts.
\label{dndme}}
\end{figure}

\placefigure{dndme}

For ``mass-conserving'' density evolution (shown in Figure 3 as a
solid line) we require a fairly high rate of density-evolution,
$(1+z)^Q$, where $Q=4-4.5$ and a milder rate of luminosity evolution
$(1+z)^b$ than above where $b=0.2-1.2$ in order to approximately
reproduce both the number counts and the angular sizes.  With the
specified large merging rate and a consequently larger number density
at moderately high redshifts ($z > 1$), it is trivial to reproduce the
observed number of galaxies in the HDF.  Furthermore, with suitable
adjustments of the relative amounts of luminosity and density
evolution ($Q/b$), it is possible to simultaneously produce a rough
fit to the angular size distributions as well for any choice of
$\Omega$ (Figure 4).

\subsection{Angular Sizes}

We compare the distributions of half-light radii recovered from the
HDF with those of our simulations in Figure 4.  Clearly, at bright
magnitudes ($21 < I_{814,AB} < 22.6$), the angular sizes recovered
from the simulations agree well with the observations as expected
given the selection of our input prototypes from this same magnitude
range.  However, at fainter magnitudes, the half-light radii from the
no-evolution simulations (hatched area in Figure 4 indicating the 1
$\sigma$ uncertainty based on the size of the bright sample) are
significantly larger than for the observations being somewhat reduced
for the cases of lower $\Omega$ (see BBS-I).  Adding our ad-hoc dwarf
population (shown in Figure 4 as a dotted line) to the no-evolution
results, we still recover $3.6 \pm 1.2$, $4.5 \pm 1.3$, and $8.1 \pm
2.2$ times fewer objects than we recover in the HDF for the size
interval 0.15 arcsec $< r_{hl} <$ 0.2 arcsec and magnitude interval
$24 < I_{814,AB} < 26$ for $\Lambda = 0.9$/$\Omega = 0.1$, $\Omega =
0.1$, and $\Omega = 1.0$, respectively.

\begin{figure}
\epsscale{1.00}
\plotone{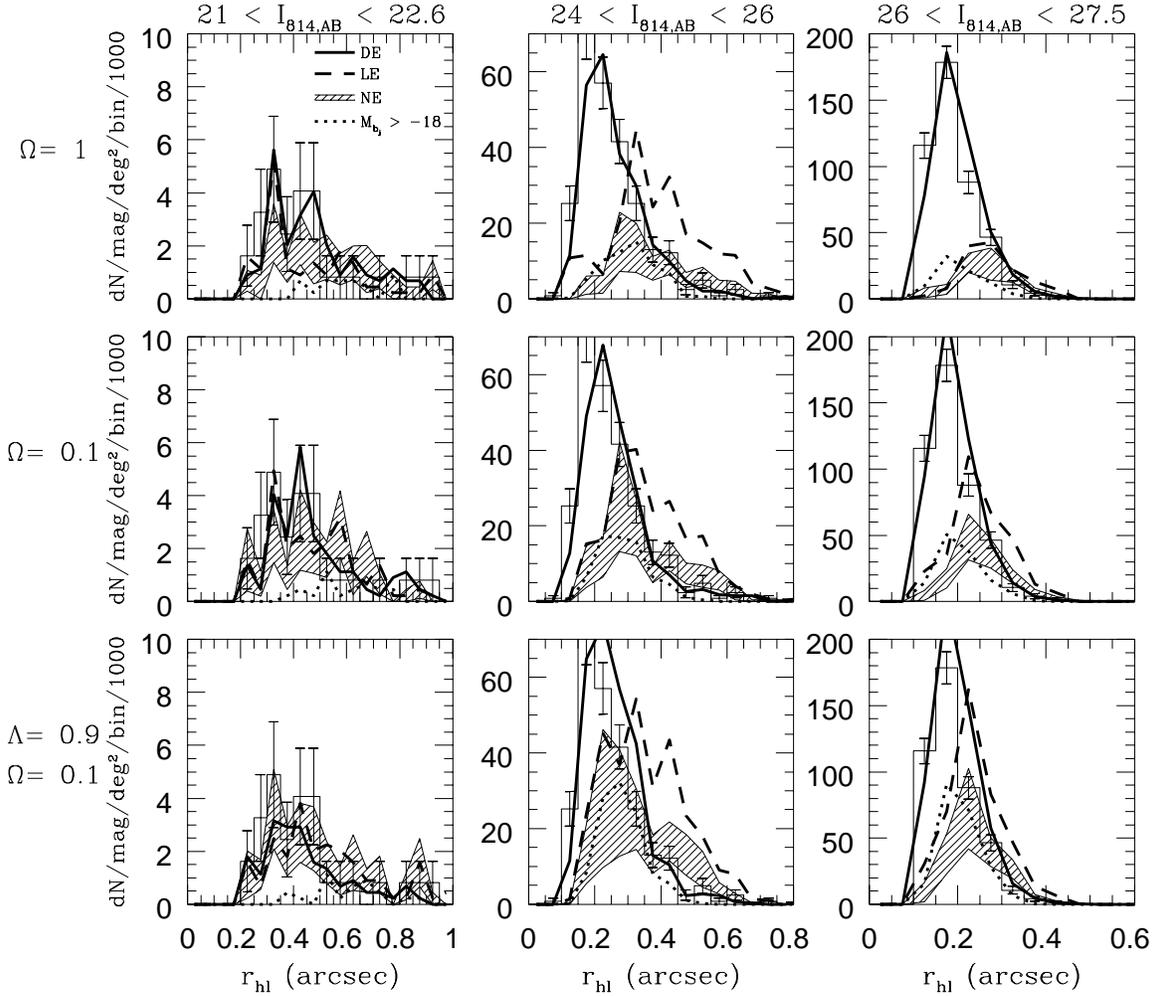}
\caption{A comparison of the observed distribution of half-light radii
recovered from the HDF (histogram with $1 \sigma$ Poissonian
uncertainties) with the distribution of half-light radii recovered
from pure luminosity evolution simulations (long dashed line) and
density evolution simulations (solid line).  For reference, we have
included the angular size distribution of no-evolution simulations
(hatched region with 1 $\sigma$ uncertainties) and our maximal dwarf
model (dotted curve).  Clearly, the angular sizes for our luminosity
evolution simulations are too large even for low $\Omega$.  In
contrast, the angular sizes for our merger models provide a rough
match to the angular sizes recovered from the data.\label{angdistae}}
\end{figure}

\placefigure{angdistae}

Incorporating our pure luminosity prescription (shown in Figure 4 as a
dashed line) only worsens the situation.  Not only are the recovered
angular size distributions much larger than those observed, but also
the number of large galaxies recovered is clearly in excess of the
data.  This significant shift to larger sizes results from the general
shift to higher redshifts and hence lower surface brightness at fixed
magnitude where the $(1+z)^4$ cosmological dimming wins over the
$(1+z)^{1.4-2.5}$ surface brightness evolution required to enhance the
predicted counts.

Therefore, unless we have greatly erred in constraining the properties
of the dwarf population, it seems clear that a large fraction of faint
galaxies are intrinsically smaller than our low-redshift input galaxy
sample.  Accordingly, it is not surprising that our mass-conserving
density evolution prescription is quite successful in allowing us to
match the observed sizes, while at the same time allowing us to easily
match the total number of galaxies observed to a given magnitude limit
(Figure 3).  For this two-parameter model, the range of $Q$ and $B$
which fit both the counts and the sizes is moderately well-constrained
by the data.  The counts are most sensitive to $B$, and the sizes to
$Q$, allowing us some independence in deriving the rates.

\subsection {Completeness and Overcounting}

Since we can easily match up our input generated catalogue with the
recovered properties of galaxies (see BBS-I), it is simple to
determine quantities like the incompleteness.  Figure 5 shows that the
incompleteness becomes significant in the range $I_{814,AB} > 26$ for
both our no-evolution simulations and those based on our simple
luminosity evolution prescriptions.  Incompleteness results from the
fact that at increasingly faint magnitudes, detection requires smaller
and hence intrinsically higher surface brightness galaxies at fixed
magnitude.  Consequently the merger and dwarf models suffer less
incompleteness at a given magnitude because of their inherently
smaller sizes.  Of course, evaluating the incompleteness is ultimately
model-dependent, but the gentle rollover at faint magnitudes occurs in
a very similar way for our merger model as in the HDF, consistent with
our finding above that the faint galaxies have small intrinsic angular
sizes.

\begin{figure}
\epsscale{1.05}
\plotone{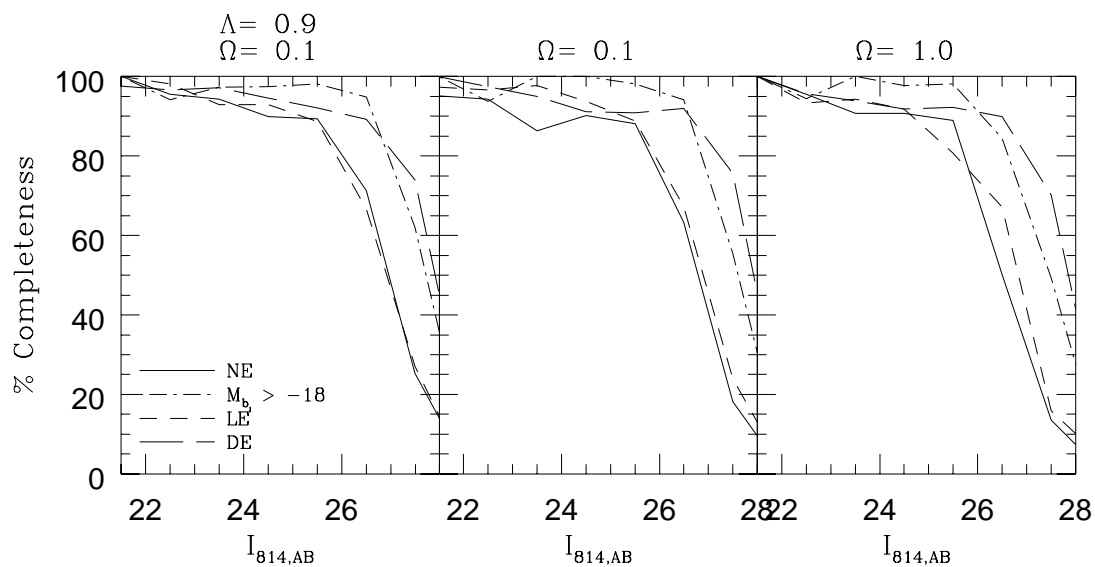}
\caption{The completeness of the $I_{814,AB}$-band counts determined
from the simulations for the no-evolution simulations (solid line),
the luminosity evolution simulations (dashed line), and the density
evolution simulations (dot-dashed line).  All models are shown for
$\Omega = 0.1$/$\Lambda = 0.9$, $\Omega=0.1$, and $\Omega=1$.  Because
surface brightness has a rough inverse proportionality to angular size
at a given magnitude, the completeness limit is directly related to
the angular sizes of the faint galaxy population in that galaxy
populations with smaller angular sizes are more complete at fainter
magnitudes.\label{completee}}
\end{figure}

\placefigure{completee}

In a similar manner, we determine the rate at which we overcount the
galaxy population in our simulated fields since each image detected
can be traced back to only one galaxy in the input catalogue.  In
Figure 6, we display the rate of overcounting for all the simulations
performed.  Evidently, in our simulations, overcounting is never an
important effect.  The worst case is for luminosity evolution,
especially the $(1+z)^{2.5}$ brightening rate used in the $\Omega=1$
geometry.  With this form of evolution the redshift distribution
extends to high redshifts ($z \sim 2-5$), where the bright ultraviolet
light of the HII regions clearly stands.  Several prominent examples
of these galaxies are evident in the simulated images (see Figure 1).

\begin{figure}
\epsscale{1.05}
\plotone{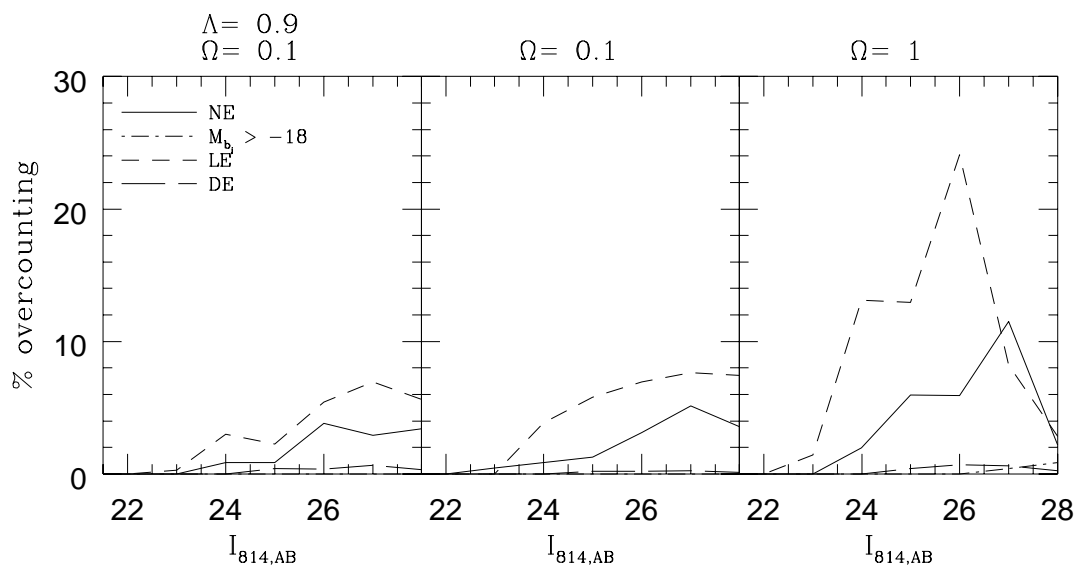}
\caption{The \% of galaxies which are counted more than once, mostly
as a result of the fact that in the UV they break up into distinct
pieces.  The overcounting rate is relatively low and similar for all
models, except for the $\Omega=1$ luminosity evolution prescription
where $L \propto (1+z)^{2.5}$.  For this model, the overcounting rate
is much greater because a large percentage of galaxies in any faint
magnitude bin at higher redshifts ($z\sim2-5$, see Figure 7) break-up
into distinct lumps due to the strong differential k-correction over
the surface of the galaxy (see Figure 1 for a dramatic illustration
of this).\label{overcounte}}
\end{figure}

\placefigure{overcounte}

\subsection{Redshift Distributions}

In Figure 7, we plot the predicted redshift distributions of the
galaxies recovered by matching up those objects recovered by
SExtractor with our input catalogues.  Without evolution, very few
galaxies lie beyond a redshift $z=2$, even at the faintest magnitudes.
Similarly, for our density-evolution models, galaxies also have rather
low redshifts, a direct consequence of the increasingly small sizes
and luminosities of galaxies in this prescription.  In contrast,
luminosity evolution accesses much higher redshifts as individual
objects are enhanced in luminosity.

\begin{figure}
\epsscale{1.00}
\plotone{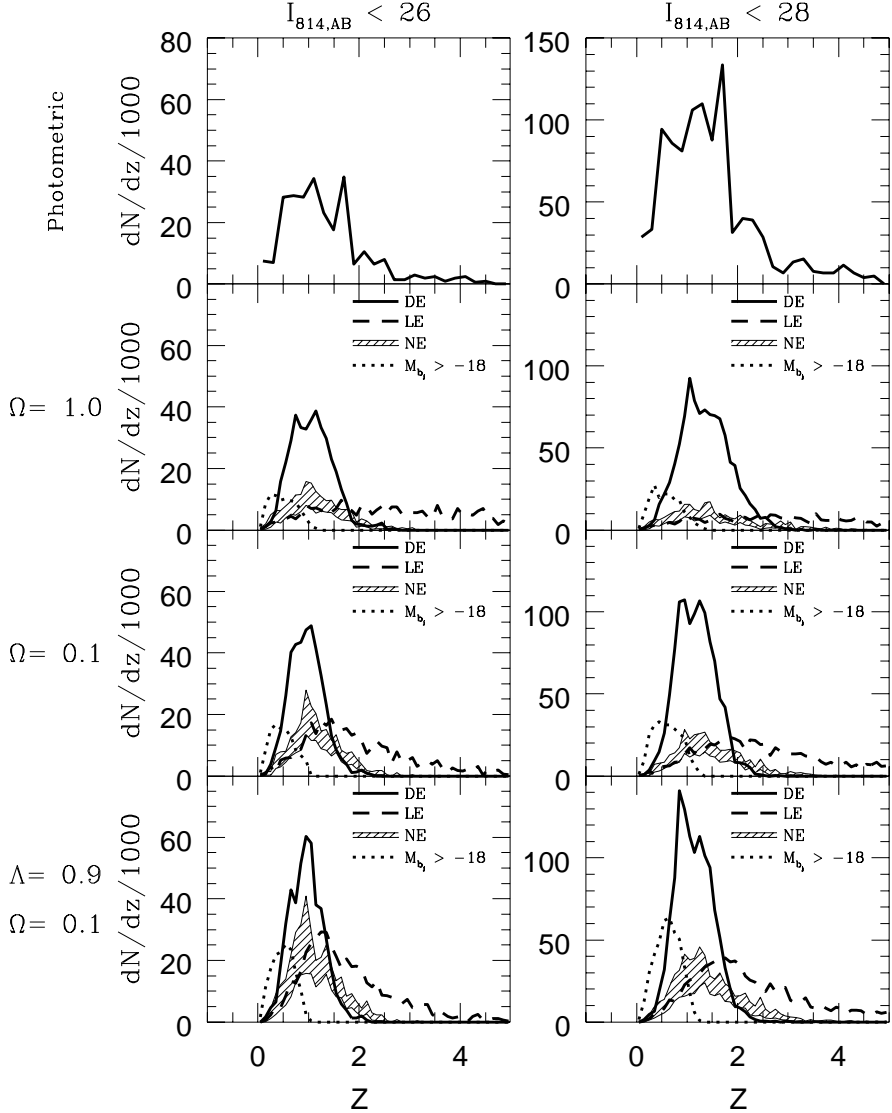}
\caption{Redshift distribution of those objects recovered by
SExtractor from our no-evolution simulations (hatched region indicated
$1 \sigma$ uncertainties), our luminosity evolution simulations
(dashed line), our density evolution simulations (solid line), and our
maximal dwarf model (dotted line) with $I_{F814W,AB} < 26$ and
$I_{F814W,AB} < 28$ for $\Omega = 0.1$/$\Lambda = 0.9$, $\Omega =
0.1$, and $\Omega = 1$ geometries.  The luminosity evolution model
produces a long tail to high redshift whereas the no-evolution
simulations and density-evolution simulations have few galaxies above
$z = 2$.  For comparison, the upper panel shows the redshift estimates
by Lanzetta, Yahil, \& Fernandez-Soto (1996) which are lower than the
estimates by Mobasher et al.\ (1996) and higher than those of Sawicki,
Lin, \& Yee (1997).\label{obsze}}
\end{figure}

\placefigure{obsze}

\subsection{$U_{300}$ and $B_{450}$ ``Dropouts''}

As in BBS-I, we can compare the number of high-redshift dropout
galaxies recovered from our simulations with those found in the HDF.
We present the Madau et al.\ (1996) colour-colour criterion for the
$U_{300}$ and $B_{450}$ dropouts in Figures 8-9 and tabulate the
number identified in Table 3.

\begin{figure}
\epsscale{0.80}
\plotone{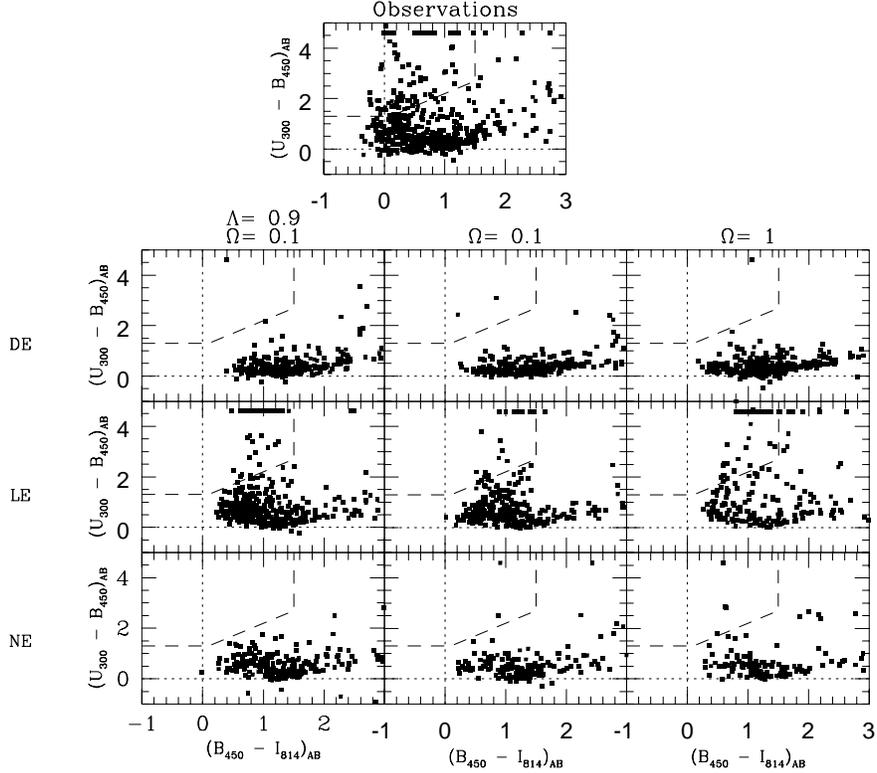}
\caption{Comparison of the $(U_{300} - B_{450})_{AB}$ versus $(B_{450}
- I_{814})_{AB}$ diagrams for our no-evolution simulations (lowest
panel), our luminosity evolution simulations (second highest panel),
and our density evolution simulations (top panel) with the
observations for $B_{450,AB} < 26.9$ (the same criterion used in Madau
et al.\ 1996).  The area interior to the dashed line is the region
Madau et al.\ (1996) suggests is occupied by high-redshift galaxies
($2<z<3.5$) whose Lyman limit crosses the $U_{300}$ passband.  Note
that galaxies near the top of the colour-colour diagram, i.e., with
$(U_{300} - B_{450})_{AB} > 4.6$, are simply lower limits on the
$(U_{300} - B_{450})_{AB}$ colour.  Both the no-evolution simulations
and our merging prescriptions underpredict the number of dropouts in
this region though the luminosity evolution models seem to do much
better in matching these observations.
\label{scue}}
\end{figure}

\begin{figure}
\epsscale{0.8}
\plotone{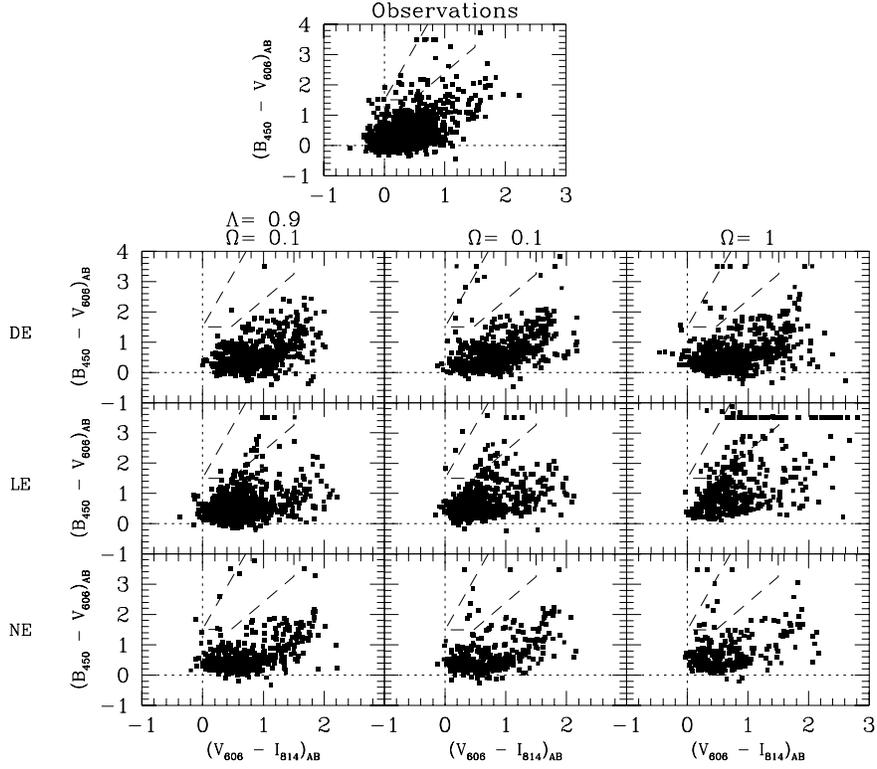}
\caption{Comparison of the $(B_{450} - V_{606})_{AB}$ versus $(V_{606}
- I_{814})_{AB}$ diagrams for our no-evolution simulations (lowest
panel), our luminosity evolution simulations (second highest panel),
and our density evolution simulations (top panel) against the
observations for galaxies with $V_{606,AB} < 28.0$ (the same criterion
used in Madau et al.\ 1996).  With a dashed line, we have overplotted
the $B$-band dropout region suggested by Madau et al.\ (1996) for
finding high redshift ($3.5 < z < 4.5$) whose Lyman limit crosses the
$B$ bandpass.  Note that galaxies near the top of the colour-colour
diagram, i.e., with $(B_{450} - V_{606})_{AB} > 3.5$, are simply lower
limits on the $(B_{450} - V_{606})_{AB}$ colour.  Similar conclusions
hold as for the $U_{300}$ dropouts.\label{sce}}
\end{figure}

\placefigure{scue}
\placefigure{sce}

\placetable{dropouts}

As in our no-evolution simulations (see BBS-I), our simple merging
prescription underpredicts the numbers of $U_{300}$ and $B_{450}$ band
dropouts.  Of course, with luminosity evolution, the numbers of
dropouts are higher.  Clearly, a more realistic inclusion of the star
formation activity in these models and the subsequent shift of the
bolometric flux into the ultraviolet would further serve to increase
the number of high redshift objects in both the pure luminosity and
density evolution models.  Taking this into account, we might expect
to find an excess of dropouts in our luminosity evolution model,
similar to the findings of Ferguson \& Babul (1998) and Pozzetti et
al.\ (1998), as well as a larger number of dropouts in our density
evolution model.

\section{Advanced Camera}

We can use our simulations to predict the likely quality of even
deeper images to be obtained using the HST Advanced Camera.  Its Wide
Field Camera (WFC) promises to have a throughput which is $\sim3.5$
greater than the peak thoughtput of WFPC2 at $\sim6500$ $\AA$, $\sim5$
greater in the $B$ band ($\sim4500$ $\AA$), and $\sim7$ greater in the
$Z$ band ($\sim9000$ $\AA$).  Figure 2 shows a direct comparison of
the image quality of WFPC2 (HDF depth) with that obtainable with the
AC (simply using the improved response) where our preferred $\Omega=1$
density evolution model is used for the simulation.  In addition to
the nominal $0.^m7$ to $1.^m1$ improvement in depth due to the
increased sensitivity, the field of view of the WFC will be 200'' x
204'', twice as large as WFPC2, with a pixel scale half as small so
that by drizzling (not included here) further improvements in
resolution and therefore depth (for the smallest objects) are
expected.

In addition, since we are currently limited by the lack of good UV
images, the High Resolution Camera (HRC) of the Advanced Camera,
though of more limited area coverage (26'' x 29''), will in principle
provide a much large sample of UV-optical imaged galaxies from which
to expand an input prototype sample, such as was used in the present
work.  The UV performace is 10 times higher than WFPC2 (Ford et al.\
1997) at $2000\AA$ and extends from $2000\AA$ to $10000\AA$.  In light
of this superior performance from the UV to the optical, the Advanced
Camera should furthermore be quite powerful in identifying a large
drop-out population up to and through the $V$ band.

\section{Discussion}

Our most interesting finding here relates to the sizes of the faint
images.  The sizes are smaller in projected areas than our
no-evolution extrapolation of the input bright galaxy sample
($\bar{z}\sim0.5$).  Adding in simple luminosity evolution only makes
the situation worse, particularly for high $\Omega$ as sizes then
effectively become larger than in the NE model.  This, together with
the lack of bright blue E/SO galaxies, make it hard to accept the
traditional view that the observed evolution is merely dominated by
the stellar evolution.  It is easier to accommodate the size evolution
and count excess by dropping the usual assumption of space-density
conservation and replacing it with the more general and physically
motivated idea that mass is conserved.  A realistic merger model,
involving star formation induced during gas-rich mergers is needed to
fully develop this approach.  For simplicity, we have chosen to
present a cruder model that nevertheless reveals the general effect.
Here we have traded image size for space density, so that the product
is independent of redshift, which for disk galaxies approximates mass
conservation.  Although we underpredict the fraction of dropouts, a
more realistic attempt to mock up this form of evolution would include
the enhancement in luminosity associated with the starburst phase
which is well documented from local examples of merging and
interacting galaxies (e.g. Joseph et al.\ 1984).  A self-consistent
modification of this model can be imagined which might draw on the
observed evolution of the blue starburst population observed in the
field and its continuation to high redshift found by Cowie et al.\
(1996) and Steidel et al.\ (1996).

Contrary to our conclusion regarding the HDF, Ferguson \& Babul (1998)
find in their modeling of luminosity evolution that the size
distribution of the faint galaxy population observed is consistent
with a low $\Omega$ Universe. Ferguson \& Babul (1998) do not evolve
the physical sizes of their galaxies and base their input parameters
on local galaxy observations.  Whilst we are not completely sure of
the source of this disagreement, we suspect that the surface
brightnesses and normalizations they use for the lowest luminosity
objects may be too high relative to observations (cf. McGaugh \& de
Blok 1997).  In any case, we are inclined to place greater weight on
our finding because of our direct use of real two-dimensional images
for a complete sample of galaxies.

\section{Conclusions}

Pure evolution in luminosity and hence in surface brightness can be
made to match the number-count excess for geometries with large volume
elements (i.e. $\Lambda=0.9$/$\Omega=0.1$).  Unfortunately, such
models reveal a discrepancy in terms of the angular sizes of the faint
galaxy population.  Clearly, galaxies would seem to evolve such that
their sizes become intrinsically smaller as well as more numerous.
``Mass-conserving'' density evolution (merging) can easily be made to
achieve this goal, but at the expense of lowering the mean redshift
and thereby underpredicting the observed ``dropout'' rate in the
$U_{300}$ and $B_{450}$ bands.  This problem might be somewhat
alleviated by accounting for the expected shift in bolometric
luminosity during the merger of gas rich systems.  Consequently, it
might be worth pursuing in more detail the effect of starburst
activity on the appearance and pixel-by-pixel spectral energy
distribution of the sample in question.  Dwarfs can also be added, but
the low-redshift constraint on their frequency is such that they would
have to have an extremely steep slope or rapid evolution, neither of
which appear supported by current observations.

\acknowledgements

We would like to thank Daniela Calzetti, Marc Davis, Mike Fall,
Holland Ford, Andy Fruchter, Nick Kaiser, Piero Rosati, and Alex
Szalay for some very useful conversations, Emmanuel Bertin for
answering several of our questions regarding SExtractor, Gordon
Squires for the use of several routines from his software package
IMCAT, Stephane Charlot for his continued helpfulness with regard to
questions we had about his spectral synthesis tables, Harry Ferguson
for his help in producing colour images and answering a few of our
questions, and finally Harry Ferguson, Steve Zepf, Eric Gawiser, and
Jonathan Tan for some helpful comments on near-final drafts of this
document.  RJB acknowledges support from an NSF graduate fellowship,
TJB acknowleges the NASA grant GO-05993.01-94A, and JS acknowledges
support from NSF and NASA grants.

{}

\newpage

\newpage

\newpage

\newpage

\newpage

\newpage

\clearpage
\newpage

\newpage

\newpage

\begin{deluxetable}{ccccc}
\tablewidth{0pt}
\tablecaption{Parameterizations Used For Evolutionary Models given
in this work.\label{LMparam}}
\tablehead{ \colhead{Evolution} & 
\colhead{$\Omega$} & \colhead{$\Lambda$} & \colhead{$B$\tablenotemark{a}} & 
\colhead{$Q$\tablenotemark{b}}}
\startdata
LE & 0.1 & 0.9 & 1.4 & 0 \\
LE & 0.1 & 0 & 1.5 & 0 \\
LE & 1.0 & 0 & 2.5 & 0 \\
DE & 0.1 & 0.9 & 0.2 & 4.0 \\
DE & 0.1 & 0 & 0.5 & 4.5 \\
DE & 1.0 & 0 & 1.2 & 4.3 \\
\enddata
\tablenotetext{a}{Surface Brightness ($\mu$) $\propto (1+z)^{B}$}
\tablenotetext{b}{Number Density $\propto (1+z)^{Q}$}
\end{deluxetable}

\begin{deluxetable}{cccccccc}
\tablewidth{0pt}
\tablecaption{Model parameters for our estimated low-luminosity 
 galaxy population.\label{lowl}}
\tablehead{
\colhead{$\Omega$} &
\colhead{$\phi_o$} & \colhead{$M^*_{b_j}$} &
\colhead{$\alpha$} & \colhead{$\tau$} &
\colhead{IMF} & \colhead{$t_f$ (Gyr)} &
\colhead{$\mu_0 ^{b_J}$\tablenotemark{a}}}
\startdata
0.1 & 1.73 & -21.14 & -1.24 & $\tau_{10}$\tablenotemark{b}
& Scalo & 16 & 22.75\\
  & 0.81 & -21.14 & -1.24 & cons\tablenotemark{c} & Salpeter & 16 & 22.75\\
\\
1 & 1.73 & -21.14 & -1.24 & $\tau_8$\tablenotemark{b} & Scalo & 12.7 & 22.75\\
  & 0.81 & -21.14 & -1.24 & cons & Salpeter & 12.7 & 22.75\\
\enddata
\tablenotetext{a}{Central surface brightness (A0V magnitudes)}
\tablenotetext{b}{Exponential SFR characterized by decay times $\tau
_{10}=10$ Gyr and $\tau_8=8$ Gyr}
\tablenotetext{c}{Constant SFR}

\end{deluxetable}

\begin{deluxetable}{ccc}
\tablewidth{0pt}
\tablecaption{Number of $U_{300}$ and $B_{450}$ dropouts.  One $\sigma$
uncertainties are given on all simulated results based on the finite
size of our bright sample.\label{dropouts}}
\tablehead{ \colhead{Data set} & \colhead{$U_{300}$ dropouts} & 
\colhead{${B_{450}}$ dropouts}}
\startdata
Observations (Madau et al.\ 1996) & 58 & 14 \\
Observations (This work) & 90 & 19 \\
NE ($\Omega = 0.1$/$\Lambda = 0.9$) & $1 \pm 1$ & $ 3 \pm 2$ \\
NE ($\Omega = 0.1$) & $2 \pm 1$ & $4 \pm 3$ \\
NE ($\Omega = 1$) & $6 \pm 3$ & $5 \pm 3$ \\
LE ($\Omega = 0.1$/$\Lambda = 0.9$) & $35 \pm 12$ & $9 \pm 3$ \\
LE ($\Omega = 0.1$) & $21 \pm 8$ & $17 \pm 6$ \\
LE ($\Omega = 1$) & $55 \pm 18$ & $33 \pm 10$ \\
DE ($\Omega = 0.1$/$\Lambda = 0.9$) & $1 \pm 1$ & $1 \pm 1$ \\
DE ($\Omega = 0.1$) & $1 \pm 1$ & $2 \pm 1$ \\
DE ($\Omega = 1$) & $2 \pm 2$ & $2 \pm 2$ \\

\enddata
\end{deluxetable}

\end{document}